# Optical interface states protected by synthetic Weyl points

Qiang Wang(王强) [1†], Meng Xiao(肖孟) [2,3†], Hui Liu(刘辉) [1*],

Shining Zhu(祝世宁) [1], and C. T. Chan(陈子亭) [2*]

[1]*National Laboratory of Solid State Microstructures, School of Physics, Collaborative Innovation Center of Advanced Microstructures, Nanjing University, Nanjing 210093, China*
[2]*Department of Physics and the Institute for Advanced Study, the Hong Kong University of Science and Technology, Clear Water Bay, Kowloon, Hong Kong*
[3]*Department of Electrical Engineering, and Ginzton Laboratory, Stanford University, Stanford, California 94305, USA*

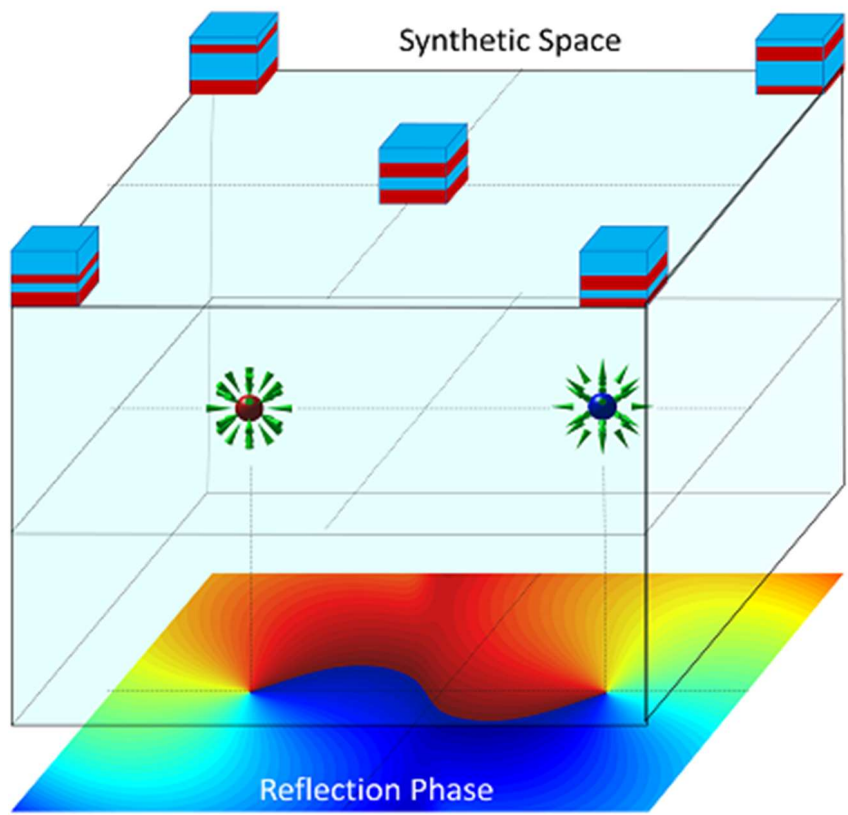

Solids can traditionally be classified as metals, semiconductors, and semimetals. A Weyl semimetal is a special kind of semimetal with unusual properties, such as always behaving as a semimetal even if perturbed. At the heart of Weyl semimetals are Weyl points, where energy bands touch at a nodal point. Important as they are, their experimental realizations are based on complex three-dimensional structures that are difficult to make, which limits the exploration of their properties. Recent theoretical work suggested that synthetic (man-made) dimensions allow for flexible control over system parameters and can hence facilitate observations of many novel phenomena. Here, we show that Weyl point physics can be explored easily using the concept of synthetic dimensions.

In particular, we consider topological nodal points in a mixed space of momentum and real space structural parameters. Such synthetic nodal points enable us to study Weyl point physics and their topological consequences in simple layer-by-layer structures, which are much easier to fabricate and characterize than ordinary Weyl crystals. As such, it enables the experimental investigation of Weyl physics in the optical regime, which is otherwise very challenging to realize. In addition, the existence of surface states in one-dimensional photonic crystals with complex unit cells can now be understood as topological consequences of nodal points in a higher-dimensional synthetic space.

Looking ahead, our approach not only increases the flexibility of realizing topological physics, but it also provides the possibility of manipulating topological matter in real time.

# Optical interface states protected by synthetic Weyl points

Qiang Wang,[1†] Meng Xiao,[2,3†] Hui Liu,[1*] Shining Zhu,[1] and C. T. Chan[2*]

[1]*National Laboratory of Solid State Microstructures, School of Physics, Collaborative Innovation Center of Advanced Microstructures, Nanjing University, Nanjing 210093, China*
[2]*Department of Physics and the Institute for Advanced Study, the Hong Kong University of Science and Technology, Clear Water Bay, Kowloon, Hong Kong*
[3]*Department of Electrical Engineering, and Ginzton Laboratory, Stanford University, Stanford, California 94305, USA*

† These authors contributed equally. Correspondence and requests for materials should be addressed to C. T. Chan (email: phchan@ust.hk ) or to H. Liu (email: liuhui@nju.edu.cn).

Weyl fermions have not been found in nature as elementary particles, but they emerge as nodal points in the band structure of electronic and classical wave crystals. Novel phenomena such as Fermi arcs and chiral anomaly have fueled the interest in these topological points which are frequently perceived as monopoles in momentum space. Here we report the experimental observation of generalized optical Weyl points inside the parameter space of a photonic crystal with a specially designed four-layer unit cell. The reflection at the surface of a truncated photonic crystal exhibits phase vortexes due to the synthetic Weyl points, which in turn guarantees the existence of interface states between photonic crystals and any reflecting substrates. The reflection phase vortexes have been confirmed for the first time in our experiments which serve as an experimental signature of the generalized Weyl points. The existence of these interface states is protected by the topological properties of the Weyl points and the trajectories of these states in the parameter space resembles those of Weyl semimetal "Fermi arcs surface states" in momentum space. Tracing the origin of interface states to the topological character of the parameter space paves the way for a rational design of strongly localized states with enhanced local field.

# I. Introduction

Great efforts have been devoted to investigate various intriguing phenomena associated with Weyl points[1-6], such as the Fermi arc surface states[7,8] and the chiral anomaly[2] associated with electronic systems. Besides electronic systems, Weyl points have also been found in photonic [9-14], acoustic[15,16] and plasmonic [17] systems. Up to now, Weyl points have mostly been identified as momentum space magnetic monopoles, i.e. sources or sinks of Berry curvature defined in the momentum space. As such, Weyl points are usually perceived as topological nodal points in the 3D momentum space defined by Bloch momentum coordinates $k_x$, $k_y$ and $k_z$. On the other hand, a few recent works considered the topological singular points in the synthetic dimensions[18-20] instead of the momentum space. The interest in considering synthetic dimensions is fueled by the ability of realizing physics in higher dimensions [18,21-24] and the possibility of simplifying experimental designs[25]. Moreover, the possible control over the synthetic dimensions enables the experimental verification of the nontrivial topology of any closed surface enclosing the topological singular points[18,19]. Here we experimentally realize generalized Weyl points in the optical frequency regime with one-dimensional (1D) photonic crystals (PCs) utilizing the concept of synthetic dimensions. Different from previous works [18-20] that replaced all three dimensions with synthetic dimensions, here we replace two wave vector components with two independent geometric parameters (which form a parameter space) in the Weyl Hamiltonian and keep one dimension as the wave vector. By doing this, we retain the novel bulk-edge correspondence relation between the edge states and the Weyl points. This is not possible for other topological singular points in purely synthetic dimensions [18-20]. Meanwhile, our hybridized Weyl points also preserve the advantage of using synthetic dimensions which facilitates the experimental investigation of Weyl physics in the optical region, whose structures are otherwise complex[11,12,15,26,27] and hence difficult to fabricate at such frequencies.

The Weyl Hamiltonian can be written as $H = \sum_{i,j} k_i v_{i,j} \sigma_j$ [28], where $v_{i,j}$, $k_i$ and $\sigma_i$ with $i = x, y, z$ represent the group velocities, wave vectors and the Pauli matrices, respectively. Unlike two-dimensional (2D) Dirac points, Weyl points are stable against perturbations when the wave vectors remain as good quantum numbers and the Weyl points do not interact with other bands[9]. Each Weyl point has its associated topological charge given by the Chern number of a closed surface enclosing it[5]. Though Weyl points are usually defined in the 3D momentum space, very recently Weyl physics has been theoretically discussed in synthetic dimensions[25]. Instead of using all three components of the wave vector, here we replace two wave vector components with two independent geometric parameters, and experimentally investigate the generalized Weyl points with simple 1D PCs. Such generalizations preserve the standard Weyl point characteristics such as the associated topological charges [11,13] and robustness against variations in the parameters [9].

Moreover, optical interface states can be found between the PCs possessing synthetic Weyl points and any reflection substrates, whose existence is stable and protected by the topological properties of the synthetic Weyl points. We show that the reflection phase of a truncated PC exhibits vortex structures [29] in the parameter space around a synthetic Weyl point. These vortexes carry the same topological charges as the corresponding Weyl points. The reflection phase along any loop in the parameter space enclosing the center of the vortex (also the position of the Weyl point) varies continuously from $-\pi$ to $\pi$. This property guarantees the existence of interface states at the boundary separating the PC and a gapped material such as a reflecting substrate [30,31] independent of the properties of the substrate. The above physical interpretation severs as the bulk-edge correspondence [30,32] for the synthetic Weyl points in our system. These interface states can be regarded as analogues of edge states in Weyl semimetals [3,6,7], and they can be useful in nonlinear optics[33-35], quantum optics [36,37], thermal radiation[38] and so on[39-42]. The winding of the reflection phase has also been discussed in the context of adiabatic charge pumping and Floquet Weyl phases in a three-dimensional network [43-45]. Furthermore, we also introduce a third geometric parameter, which extends the three dimensional space to four dimensions. By tuning the third geometric parameter, we also observe the topological transition from Weyl semimetals to nodal line semimetals[9,46].

## II. Results

### A. Synthetic Weyl points in parameter space

To illustrate the idea of synthetic Weyl points in a generalized parameter space, we consider a 1D PC consisting of four layers per unit cell, as shown in the inset in Fig. 1(a). In our experiments, the first and third layers (blue) are made of $H_fO_2$ with refractive index $n_a$ =2.00, and the second and fourth layers (red) are made of $S_iO_2$ with refractive index $n_b$=1.45. The thickness of each layer is given by:

$$\begin{aligned} d_{a1} &= (1+p)d_a \\ d_{b1} &= (1+q)d_b \\ d_{a2} &= (1-p)d_a \\ d_{b2} &= (1-q)d_b \end{aligned} \quad (1)$$

Since the thickness of each layer cannot be a negative value, so *p* and *q* are both fall in [-1,1], which makes the p-q space is an closed parameter space. The total optical length $L$ inside the unit cell is a constant $2(n_a d_a + n_b d_b)$ for the whole $p$-$q$ space. As illustrated in Fig. 1(a), the structural parameters $p$ and $q$, together with one Bloch wave vector $k$, form a 3D parameter space in which Weyl physics can be studied.

We start with the PCs with only two layers inside each unit cell, a layer of H$_f$O$_2$ with thickness $d_a$ and a layer of S$_i$O$_2$ with thickness $d_b$. The band dispersion is plotted in Fig. 1(b) in red. A four-layer PC with parameters $p=0$ and $q=0$ simply doubles the length of each unit cell and folds the Brillouin zone. The dispersion of this four-layer PC is shown in Fig. 1(b) in blue. Such artificial band folding gives a linear crossing along the wavevector direction. Away from the point where $p=0$ and $q=0$, the degeneracy introduced by the band folding is lifted and a band gap emerges. Fig. 1(c) shows the band dispersions in the $p$-$q$ space with $k=0.5k_0$, where $k_0 = \pi/(d_a + d_b)$. Two bands form a conical intersection indicating that band dispersion is linear in all directions. To characterize this degenerate point, we derive an effective Hamiltonian for parameters around it (see Appendix A):

$$H = p v_{pz}\sigma_z + q v_{qy}\sigma_y + \xi_k v_{kx}\sigma_x, \qquad (2)$$

where $\xi_k = (k - 0.5k_0)/k_0$, $v_{pz} = 0.1073$, $v_{qy} = -0.0946$ and $v_{kx} = -1.985$ for our system. This Hamiltonian possesses a standard Weyl Hamiltonian form and with the Weyl node located at $(p, q, k) = (0, 0, 0.5k_0)$ carrying a "charge" of -1 according to the usual definition. The topological charge of this Weyl point can also be numerically verified, using the method in ref 5 to calculate the Weyl point charge. As shown in the inset in Fig. 1(d), the Berry phases are defined on a spherical surface with a fixed azimuthal angle $\theta$ (red circle). We then track the evolution of the Berry phases as a function of $\theta$, as shown in Fig. 1(d), which shows that the Chern number of the band below the Weyl point decreases by one as $\theta$ increases from 0 to $\pi$. This implies that the Weyl point has a negative charge. (See Appendix B for more details).

In addition to the Weyl point constructed above, we also find Weyl points on higher bands and at different positions in the parameter space. In Fig. 2, we show all the Weyl points on the lower five bands at either $k=0.5k_0$ or $k=0$. The locations and charges of these Weyl points are marked in the insets. The Weyl point between band 1, the lowest band, and band 2 (Fig. 2(a)) with charge -1 has already been discussed above. Two Weyl points with charge -1 exist between band 2 and band 3. Weyl points with charge +1 appear on higher bands. To realize a Weyl point in the reciprocal space, one needs to break either inversion symmetry or time-reversal symmetry. Our systems possess time reversal symmetry but do not exhibit inversion symmetry for general $p$ and $q$. Time reversal symmetry requires that the charges of two Weyl points at $(p_0, q_0, k_0)$ and $(p_0, q_0, -k_0)$ are the same. Besides time reversal symmetry, our systems also exhibit the symmetry $\varepsilon(p,q,x) = \varepsilon(-p,q,-x) = \varepsilon(p,-q,-x) = \varepsilon(-p,-q,x)$, where $\varepsilon$ is the permittivity and $x$ represents the real space position. This symmetry ensures that as long as there is a Weyl point at $(p_0, q_0)$ at a particular frequency, there will be other Weyl points at $(-p_0, q_0)$, $(p_0, -q_0)$ and $(-p_0, -q_0)$, and these Weyl points will all possess the same topological charge. Note that while the total charges of Weyl points must vanish in periodic systems[47] as the reciprocal space is periodic, such a constraint does not apply here as the parameter space is not periodic. This features one of the main difference between Weyl points in synthetic dimensions and those in momentum space.

**B. Reflection phases around the Weyl points**

We then consider the reflection phase of a normal incident plane wave when the PC is semi-infinite. The working frequency of the incident wave is chosen to be the frequency of the Weyl point between band 1 and band 2. Except for the $p=q=0$ point, the working frequency is inside the band gap for all other $p$ and $q$ values. Hence the reflection coefficient can be written as $r=\exp(i\phi)$, with $\phi$ being a function of $p$ and $q$. In Fig. 3(a), we show the reflection phase in the whole $p$-$q$ space, where the truncation boundary is at the center of the first layer. The reflection phase distribution shows a vortex structure, with the Weyl point at the vortex center. The topological charge of this vortex is given by the winding number of the phase gradient[48], which is the same as that of the Weyl point. For illustration purposes, we choose a circle centered at this Weyl point as marked by the gray dashed circle in Fig. 3(a). The reflection phase along the circular loop decreases with the polar angle $\varphi$, and picks up a total change of $-2\pi$ after each circling. We note that the same circle does not enclose any Weyl points between band 2 and band 3. The reflection phase at the frequency of the Weyl points between band 2 and band 3 is shown in Fig. 3(c) (the circular loop is also marked). Now the reflection phase along the loop in Fig. 3(c) covers a range much less than 2π.

The above conclusion can also be experimentally verified. We choose five PC configurations whose locations in the $p$-$q$ space are indicated by the white circle, triangle, square, diamond and pentagram in Figs. 3(a) and 3(c). The PCs here were fabricated by electron beam evaporation on a substrate made of K9 glass. Before evaporation, we cleaned the substrate with an acidic solution. During the evaporation, the pressure in the chamber was kept below $2\times10^{-3}$ Pa, and the temperature was maintained at 90℃. The uncertainty in the thickness of each layer in the fabrication was below 10nm for all the PCs. Here we set $d_a = 97nm$, $d_b = 72nm$ and the number of unit cells to 15. The layers in these five PCs are given by $(p, q) = (0.24, 0.44)$, $(-0.11, 0.49)$, $(-0.38, 0.31)$, $(-0.45, 0.22)$ and $(0.33, -0.37)$, respectively. We used two spectrometers to measure the reflection phase of PCs. One spectrometer ranges from 880nm to 1700nm with a resolution of 0.8nm (BWTEK-BTC 261E Cooled InGaAs Array Spectrometer). The other ranges from 330nm to 1070nm with a resolution of 0.8nm (BWTEK-BTC 611E Spectrometer). The reflection phase was measured using a Fabry–Perot (FP) interference setup (See the Supplementary Material Sec. I for the more detailed setup).

The measured reflection phases across the first and second band gaps are shown in Figs. 3(b) and 3(d), respectively, where different markers correspond to different samples. Black dashed vertical lines mark the frequency of the Weyl points and gray areas represent the passband region. We also numerically calculate the reflection phases and the results are shown in Figs. 3(b) and 3(d) with colored curves. The experimental results agree reasonably well with the numerical results. The reflection phases in Fig. 3(b) are uniformly distributed and cover the whole $[-\pi \ \pi)$ range, while the reflection phases in Fig. 3(d) cover only a narrow range. Whether the reflection phase trajectories span the gap region (as in Fig. 3(b)) or not (as in Fig. 3(d)) depends on whether the loop in the parameter space encloses a Weyl node or not, given the fact that the total phase change must be $2\pi$ if the loop encircles a Weyl point, a proof can be found in Appendix C. We emphasize that the vortex structure and the charge of the vortex are independent of the position where we truncate the PCs (See Supplementary I). Though the reflection phase vortex discussed here is in the generalized parameter space, the reflection phase vortex acts as a signature of Weyl physics. See for example

Ref. [45] and the discussions in Appendix C.

**C. Fermi-arc-like interface states**

The vortex structure and the associated topological charge guarantee the existence of interface states [30,31], between the PCs with Weyl points and the reflecting substrates, regardless of the substrate properties. The existence of interface states[30] is given by

$$\phi_{PC} + \phi_{S} = 2m\pi, \quad m \in \mathbb{Z}, \tag{3}$$

where $\phi_{PC}$ and $\phi_{S}$ represent the reflection phases of the PC and the reflecting substrate, respectively. As the reflection phase on the loop encircling the Weyl point covers the whole $[-\pi, \pi)$ region, no matter what the reflection phase of the reflecting substrate is, Eq. (3) can always be satisfied for at least one polar angle. If we consider loops with different radii, then the interface states form a continuous trajectory beginning from the Weyl point. The trajectory of the interface states ends either at another Weyl point with an opposite charge or at the boundary of the parameter space. The behavior of these interface states connecting Weyl points with opposite charges in the parameter space has the same mathematical origin as that of the Fermi arc [1,2,4,7] in Weyl semimetals. There is however a crucial difference: the Fermi arc starts and ends with Weyl points in a periodic system, while the interface states in our system can connect Weyl points to the boundary of the parameter space because the total charge of the Weyl points does not vanish inside the parameter space.

As an example, we consider the Weyl points on the fourth and fifth bands as shown in Fig. 2(d).

There are a total of eight Weyl points: six with charge -1 and the remaining two with charge +1. As our system does not possess any external symmetry that can relate two Weyl points with opposite charges, the frequency of the Weyl points with a positive charge is higher than that of the Weyl points with a negative charge. To turn the Weyl points into the working frequency range of our measurements, we increase the thickness of each slab in the PCs proportionally and we now have $d_a = 0.323 um$ and $d_b = 0.240 um$. We then set the working frequency to the frequency of the Weyl point with a positive charge (303THz) and truncate the PCs at the center of the first layer. A sketch of measuring the interface states was shown in Fig. 4(a). Here the silver slab represents the silver film, blue represents $H_fO_2$ and red represents $S_iO_2$. Wave incidents from the silver film side and the interface states are localized in the interface region between the silver slab and the semi-infinite PC. The amplitude of the wave is shown schematically in yellow. In Fig. 4(b), we show the reflection phase in the parameter space, where gray areas within black dashed lines mark the regions of the bulk band. If the truncated PCs are coated with silver films, then interface states exist which satisfy Eq. (3). The reflection phase of the silver film coated is measured to be $(-0.95 \pm 0.0471)\pi$ at 303THz. White dashed lines in Fig. 4(b) show the trajectories of interface states. In addition to interface states connecting Weyl points with different charges, there are also trajectories of interface states terminating at the boundary of our parameter space. We also performed experiments to verify the results. We choose four points (positions marked with cyan triangles in Fig. 4(b)) to experimentally demonstrate the existence of the interface states, with samples consisting of 10 periods, and the thickness of layers in these four PCs are given by $(p,q) = (0.50, 0.26)$, $(0.56, 0.30)$, $(0.70, 0.38)$ and $(0.78, 0.36)$ respectively. The results are shown in Fig. 4(c) (red crosses) and match well with the numerical prediction (cyan dashed line), the more detailed setup can be found in the Supplementary Material Sec. II. We emphasize here that the existence of interface states is "robust" to the property of the reflecting substrates: we can always find trajectories of interface states that link the two Weyl points with opposite charges no matter what the reflecting substrate is (See the Supplementary Material Sec. III).

**D. "Nodal lines" in higher dimensional space**

Compared with Bloch momentum space, synthetic dimensions provide a flexible way to construct topological system in higher dimensional space, which enables the study of phenomena that occur only in higher dimensional spaces[23]. As an example, we can define another parameter $R$ as the ratio $R = n_a d_a / (n_a d_a + n_b d_b)$, which belongs to (0, 1). By varying this parameter, we can now observe topological transitions of the band structures. In Fig. 5, we show the dispersion of the fourth and fifth band gaps with $k=0$ for different values of $R$. The insets show the positions of the Weyl points (black circles) and degenerate lines (dashed lines). When $R = 0.25$ (Fig. 5(c)), 0.5 (Fig. 5(e)) and 0.75 (Fig. 5(g)), only degenerate lines exist, which act as analogues of the nodal lines in semimetals[49]. When R passes through these transition points, Weyl points will immerse as node line and reappear. In the process, a pair of Weyl point with positive and negative charge will appear or disappear simultaneously. Although the number of Weyl points will changes, the total charge remains constant.

## III.  Conclusion

We have shown the existence of Weyl points in the parameter space and their topological consequences. In the specific example of dielectric superlattices, the reflection phase of the semi-infinite multilayered PC shows a vortex structure with the same topological charge as the synthetic Weyl points defined in the parameter space. The vortex structure guarantees the existence of interface states, which can be used in various systems [33-42] (See also one example given in the Supplementary Material Sec. IV). In general, interface states may or may not exist at the boundary between a 1D PC and a reflecting substrate[30]. The Weyl points here provide a deterministic scheme to construct interface states between multi-layered PCs and reflecting substrates of arbitrary reflection phases[31]. In the past, numerical simulation is the only way to predict the optical properties of photonic crystals with complex unit cells defined by many parameters and there is no easy way to connect the bulk properties to the surface properties such as reflectance and existence of interface states. Here, we see that the topological character of the nodal points in the higher dimension space defined jointly by momentum and structural parameters actually connects the properties of the bulk to those of the surface. We emphasize that the topological notions apply to all frequencies, including high frequency gaps where the PCs cannot be described by the effective medium theory.

The geometric parameters $p$ and $q$ are fixed for each PC in this work. However, there are techniques [50-53] that can be employed to control the geometric parameters as well as the refractive index in real time. Combined with the understanding of the topological origin of the interface states, such tunability allows for the control of the interface states which may facilitate various applications. In addition, the reflection phase vortex offers a flexible way to manipulate the electromagnetic wave such as generating vortex beams and controlling the reflection direction. This work also opens up a new direction for experimentally exploring the physics in topological theory in higher dimensions[23]. With more parameters involved, we can also construct Weyl points with a higher topological charge[12,13,54].

## Appendix A: Effective Hamiltonian around the Weyl points

Here we adapt the transfer matrix method to derive the effective Hamilton around the Weyl points. Each unit cell is composed of four layers, the transfer matrix can be written as

$$\mathbf{T} = \begin{pmatrix} C & D \\ D^* & C^* \end{pmatrix} \begin{pmatrix} A & B \\ B^* & A^* \end{pmatrix} \tag{A1}$$

where

$$A = e^{ik_a d_{a1}} \left[ \cos(k_b d_{b1}) + \frac{i}{2}\left(\frac{n_b}{n_a} + \frac{n_a}{n_b}\right) \sin(k_b d_{b1}) \right] \tag{A2}$$

$$B = e^{-ik_a d_{a1}} \left[ \frac{i}{2}\left(\frac{n_b}{n_a} - \frac{n_a}{n_b}\right) \sin(k_b d_{b1}) \right] \tag{A3}$$

$$C = e^{ik_a d_{a2}} \left[ \cos(k_b d_{b2}) + \frac{i}{2}\left(\frac{n_b}{n_a} + \frac{n_a}{n_b}\right) \sin(k_b d_{b2}) \right] \tag{A4}$$

$$D = e^{-ik_a d_{a2}} \left[ \frac{i}{2}\left(\frac{n_b}{n_a} - \frac{n_a}{n_b}\right) \sin(k_b d_{b2}) \right] \tag{A5}$$

And the wave equation can be written as

$$\left[\mathbf{T} - e^{i2k(d_a + d_b)}\right] \begin{pmatrix} c^+ \\ c^- \end{pmatrix} = 0 \tag{A6}$$

where $c^+$ and $c^-$ represent the coefficients of the forward propagating and backward propagating waves inside the first layer respectively. Define the following dimensionless coefficients

$$\begin{aligned} \xi_f &= (f - f_w)/f_w \\ \xi_p &= (p - p_w) \\ \xi_q &= (q - q_w) \\ \xi_k &= (k - k_w)/k_0 \end{aligned} \tag{A7}$$

Here $(p_w, q_w, k_w)$ denotes the position of the Weyl point under consideration, and $f_w$ is the frequency of the Weyl point. The system under consideration only supports one forward propagating wave and one backward propagating wave. Hence if there is a Weyl point, the Weyl point must located at the Brillouin zone center or zone boundary. Expanding $\mathbf{T}$ with respect to $(\xi_p, \xi_q, \xi_f)$ around the Weyl point, it is easy to show that the zero order term of $\mathbf{T}$ is either 1 or -1 (depending on whether the Weyl point is at the zone boundary or zone center). Hence up to the first order of $(\xi_p, \xi_q, \xi_f)$, we have

$$\mathbf{T} = \begin{pmatrix} \pm 1 + ic_1\xi_p + ic_2\xi_q + ic_3\xi_f & d_1\xi_p + d_2\xi_q + d_3\xi_f \\ d_1^*\xi_p + d_2^*\xi_q + d_3^*\xi_f & \pm 1 - (ic_1\xi_p + ic_2\xi_q + ic_3\xi_f) \end{pmatrix} + O(\xi_p^2, \xi_q^2, \xi_f^2), \quad (A8)$$

Meanwhile, as $\det(\mathbf{T}) = 1$, one can prove that $\{c_1, c_2, c_3\} \in \mathbb{R}$. Keeping only the first order, Eq. (A6) can be written as

$$\begin{pmatrix} -(c_1\xi_p + c_2\xi_q - c_4\xi_k) & -i(d_1\xi_p + d_2\xi_q) \\ i(d_1^*\xi_p + d_2^*\xi_q) & -(c_1\xi_p + c_2\xi_q + c_4\xi_k) \end{pmatrix} \begin{pmatrix} -ic^+ \\ ic^- \end{pmatrix} = \xi_f \begin{pmatrix} c_3 & id_3 \\ -id_3^* & c_3 \end{pmatrix} \begin{pmatrix} -ic^+ \\ ic^- \end{pmatrix}, \quad (A9)$$

where $\{c_1, c_2, c_3, c_4\} \in \mathbb{R}$ and $\{d_1, d_2, d_3\} \in \mathbb{C}$. Note that now the matrix on both sides of Eq. (A9) are Hermitian matrixes. The matrix on the right hand side of Eq. (A9) can be decomposed as

$$\begin{pmatrix} c_3 & id_3 \\ -id_3^* & c_3 \end{pmatrix} = U \begin{pmatrix} c_3 - |d_3| & 0 \\ 0 & c_3 + |d_3| \end{pmatrix} U^\dagger \quad (A10)$$

where

$$U = \frac{1}{\sqrt{2}} \begin{pmatrix} -id_3/|d_3| & id_3/|d_3| \\ 1 & 1 \end{pmatrix} \quad (A11)$$

After some simple mathematics, Eq. (A9) can be written as

$$\begin{pmatrix} c_3 - |d_3| & 0 \\ 0 & c_3 + |d_3| \end{pmatrix}^{-\frac{1}{2}} U^\dagger M U \begin{pmatrix} c_3 - |d_3| & 0 \\ 0 & c_3 + |d_3| \end{pmatrix}^{-\frac{1}{2}} \begin{pmatrix} -i\tilde{c}^+ \\ i\tilde{c}^- \end{pmatrix} = \xi_f \begin{pmatrix} -i\tilde{c}^+ \\ i\tilde{c}^- \end{pmatrix}, \quad (A12)$$

where

$$M = \begin{pmatrix} -(c_1\xi_p + c_2\xi_q - c_4\xi_k) & -i(d_1\xi_p + d_2\xi_q) \\ i(d_1^*\xi_p + d_2^*\xi_q) & -(c_1\xi_p + c_2\xi_q + c_4\xi_k) \end{pmatrix}, \quad (A13)$$

and

$$\begin{pmatrix} -i\tilde{c}^+ \\ i\tilde{c}^- \end{pmatrix} = \begin{pmatrix} c_3 - |d_3| & 0 \\ 0 & c_3 + |d_3| \end{pmatrix}^{\frac{1}{2}} U^\dagger \begin{pmatrix} -ic^+ \\ ic^- \end{pmatrix}. \quad (A14)$$

Eq. (A12) can be reformulated in the Pauli matrixes form as

$$\begin{pmatrix} \xi_p & \xi_q & \xi_k \end{pmatrix} \mathbf{v} \begin{pmatrix} \sigma_0 \\ \sigma_x \\ \sigma_y \\ \sigma_z \end{pmatrix} \begin{pmatrix} -i\tilde{c}^+ \\ i\tilde{c}^- \end{pmatrix} = \xi_f \begin{pmatrix} -i\tilde{c}^+ \\ i\tilde{c}^- \end{pmatrix}, \quad (A15)$$

where $\sigma_i$ ($i=x,y,z$) is the Pauli matrix, $\sigma_0$ is a $2\times 2$ identity matrix, $\mathbf{v}$ is a $3\times 4$ real matrix. Hence the matrix on the left hand side of Eq. (A12)

$$H \equiv \begin{pmatrix} \xi_p & \xi_q & \xi_k \end{pmatrix} \mathbf{v} \begin{pmatrix} \sigma_x \\ \sigma_y \\ \sigma_z \\ \sigma_0 \end{pmatrix} \quad (A16)$$

works as an effective Hamiltonian of the system. This effective Hamiltonian processes a Weyl form.

We choose the Weyl point between band 1 and band 2 as an example. Here $(p_w, q_w, k_w) = (0, 0, 0.5 k_0)$ with $k_0 = \pi/(d_a + d_b)$. For the PC with $d_a = 97 nm$, $d_b = 72 nm$, the frequency of this Weyl point is at $f_w = 247.6 THz$. Around this Weyl point, the coefficient matrix is given by

$$\mathbf{v} = \begin{pmatrix} 0 & 0 & 0.1073 & 0 \\ 0 & -0.0946 & 0 & 0 \\ -1.985 & 0 & 0 & 0 \end{pmatrix} \quad (A17)$$

Hence according to the definition, the charge of this Weyl point is -1. For other Weyl points, there are more nonzero coefficients in Eq. (A17). To verify that the obtained effective Hamiltonian works, we plot the dispersion along the three directions $p$, $q$, $k$ in Fig. 6(a), (b), and (c), respectively. Red dashed lines are calculated using the effective Hamiltonian, while the blue circles are from the full wave simulations. They agree well near the Weyl point which shows that the effective Hamiltonian works.

# Appendix B: Numerical calculations of the topological charges

We adapt a numerical method used in the literature to determine the charges of Weyl points. We first calculate the Berry phase on the circle on a spherical surface (centered at the Weyl point) with a fixed $k$. and then trace the evolutions of the Berry phases as $k$ varying from the upper pole to the lower pole. The total change of the Berry phase corresponds to the Chern number of this spherical surface. To illustrate this idea, we choose two loops $l_i$ and $l_{i+1}$ as shown in Fig. 7(a), and assume the Berry phases of them are given by $\gamma_i$ and $\gamma_{i+1}$, respectively. The Berry phases along these loops equal to the Berry fluxes out of the upper spherical crown of these two loops. The Berry phases here are well-defined up to an uncertainty of $2m\pi$ ($m \in \mathbb{Z}$). When $l_i$ and $l_{i+1}$ are very close to each other, the Berry phases of these two loops must also be close to each other and the difference between them equals to the Berry flux coming out from the strip region between $l_i$ and $l_{i+1}$. Hence the change of the Berry phase over the whole spherical surface equals to the Berry flux out of the Weyl point.

To numerical calculate the Berry phase for each loop, we use a discretized algorithm[5]:

$$\gamma_n = -\mathrm{Im} \sum_{i=1}^{N} \ln \left( \int_0^L \varepsilon(x) u^*_{n,p_i,q_i,k}(x) u_{n,p_{i+1},q_{i+1},k}(x) dx \right) \tag{B1}$$

where $\varepsilon$ represents the permittivity, and $u_{n,p_i,q_i,k}(x)$ is the periodic-in-cell part of the eigenelectric field of a state on the $n$-th band and with parameter $p_i$, $q_i$ and $k$. Here $p_i = r\sin(\theta)\cos(i2\pi/N)$, $q_i = r\sin(\theta)\sin(i2\pi/N)$ and $N$ has been chosen to be large enough such that $\gamma_n$ converges. In Fig. 7(b), we plot Berry phases on band 1 (blue) and band 2 (red) as a function of $\theta$. In the calculation, the radius of the sphere in Fig. 7 is set to be $r=0.001$. We can see that the variation range of the Berry phase is $2\pi$ for the upper band (the red line) and $-2\pi$ for the lower band (the blue line), which means the charge of this Weyl point is -1.

# Appendix C: Reflection phases around Weyl points

In this appendix, we show how to get the reflection phase from the effective Hamiltonian obtained in Appendix A. We choose the Weyl point between the first and second bands as an example. The Hamiltonian of our system near this Weyl point is given by:

$$H = \xi_p v_{pz} \sigma_z + \xi_q v_{qy} \sigma_y + \xi_k v_{kx} \sigma_x, \tag{C1}$$

where $v_{pz} = 0.1073$, $v_{qy} = -0.0946$ and $v_{kx} = -1.985$ for the Weyl point under consideration. Now let us consider an elliptical loop circling the Weyl point in the parameter space spanned by $\xi_p$ and $\xi_q$:

$$\begin{aligned} \xi_p |v_{pz}| &= r\cos(\varphi') \\ \xi_q |v_{qy}| &= r\sin(\varphi') \end{aligned}, \tag{C2}$$

where $\varphi'$ is the polar angle in the new parameter space and $\varphi' \in [0, 2\pi)$. $\varphi'$ processes a one to one correspondence to the polar angle in the original parameter space $\varphi$ through the relation:

$$\tan(\varphi')|v_{pz}| = |v_{qy}|\tan(\varphi), \tag{C3}$$

In the following, we obtain the reflection phase as a function of $\varphi'$ and show that charge of the vortex of reflection phase is equal to the charge of the Weyl point. The elliptical loop is chosen as defined in Eq. (C2) to simplify the derivation. Here we choose the working frequency to be inside the bulk band gap, and hence $\xi_f < r$. We define $\cos\chi = (\xi_f/r)$, with $\chi \in (0, \pi)$. Substitute Eq. (C2) into Eq. (C1), we obtain two solutions of $\xi_k$ as:

$$\xi_k v_{kx} = \pm ir\sin\chi, \tag{C4}$$

As expected, the wave vector $\xi_k$ becomes purely imaginary. We assume that the direction of the incident wave is along the positive direction, and hence we require $\text{Im}(\xi_k) > 0$. The corresponding eigenstate is given by:

$$\vec{u} = \begin{pmatrix} 1 \\ -i\beta \end{pmatrix}, \tag{C5}$$

where

$$\beta = -\frac{\cos(\chi) - \cos(\varphi')}{\sin(\chi) - \sin(\varphi')} = \tan\left((\varphi' + \chi)/2\right) , \tag{C6}$$

Substitute Eq. (C5) in to Eq. (A14), we obtain the coefficients of the forward propagating and backward propagating wave as

$$\begin{pmatrix} c^+ \\ c^- \end{pmatrix} = \frac{1}{\sqrt{2}} \begin{pmatrix} e^{i\psi_1}\left[\left(c_3 - |d_3|\right)^{-1/2} + i\beta\left(c_3 + |d_3|\right)^{-1/2}\right] \\ -i\left[\left(c_3 - |d_3|\right)^{-1/2} - i\beta\left(c_3 + |d_3|\right)^{-1/2}\right] \end{pmatrix} , \tag{C7}$$

where $c_3$ and $d_3$ are the coefficients defined in Eq. (A8) and $\psi_1 = \mathrm{Arg}(d_3)$. To simplify the notation, we define

$$\psi_2 \equiv \tan^{-1}\left[\beta\left(\frac{c_3 - |d_3|}{c_3 + |d_3|}\right)^{1/2}\right] , \tag{C8}$$

and

$$A_0 = \sqrt{\frac{1}{c_3 - |d_3|} + \frac{\beta^2}{\left(c_3 + |d_3|\right)}} , \tag{C9}$$

Now Eq. (C7) can be simplified as

$$\begin{pmatrix} c^+ \\ c^- \end{pmatrix} = A_0 \sqrt{\frac{1}{2}} \begin{pmatrix} e^{i\psi_1}e^{i\psi_2} \\ -e^{i\pi/2}e^{-i\psi_2} \end{pmatrix}. \tag{C10}$$

Then the corresponding electronic field at the boundary of the PC is:

$$E_{1y} = c^+ e^{ik_a x} + c^- e^{-ik_a x} = A_0\left[\left(e^{i\psi_1}e^{i\psi_2}\right)e^{ik_a x} - \left(e^{i\pi/2}e^{-i\psi_2}\right)e^{-ik_a x}\right]/\sqrt{2} , \tag{C11}$$

where the subscript "1y" means the electric field is along the $y$ direction and inside the first layer, $k_a$ represents the wave vector in the first layer, and $x$ represents the distance from the truncated plane to the starting plane of the first layer. The corresponding magnetic field is:

$$H_{1z} = \frac{1}{\sqrt{2}z_a} A_0\left[\left(e^{i\psi_1}e^{i\psi_2}\right)e^{ik_a x} + \left(e^{i\pi/2}e^{-i\psi_2}\right)e^{-ik_a x}\right] , \tag{C12}$$

where $z_a$ is the bulk impedance of the first layer (in unit of the vacuum impedance $z_0$). So the surface impedance of the photonic crystal can be written as:

$$Z_{pc} = iz_a\left(\tan(\psi_1/2 + \psi_2 + k_a x - \pi/4)\right) , \tag{C13}$$

and the corresponding reflection phase as

$$\phi = \pi + 2\arctan\left[-z_a \tan\left(\psi_1/2 + \psi_2 + k_a x - \pi/4\right)\right]. \quad (C14)$$

To show the validity of this method, we compare the reflection phase obtained from Eq. (C14) to that from the full wave simulation (where we use the transfer matrix method). As the working frequency is inside the bandgap and if the number of unit cell is large enough, the reflection phase will converge. In the full wave simulation, we have ensured that the number of unit cells is large enough. In Fig. 8, red circles and blue solid line represent the reflection phases obtained with Eq. (C14) and the transfer matrix method, respectively. They agree quite well with each other, which shows that Eq. (C14) works in the region near the Weyl point.

Now let us continue to analyses the monotonicity and dependence of the reflection phase obtained in Eq. (C14). It is easy to see that $\phi$ is a monotonically decreasing function of $\psi_1/2 + \psi_2 + k_a x - \pi/4$. The dependences of $\phi$ on the polar angle $\varphi$ and the working frequency are all inside the phase $\psi_2$. For the Weyl point under consideration, $|c_3| > |d_3|$ and hence according to Eq. (C8), $\psi_2$ is a monotonically increasing function of $\beta$. Note $\cos\chi = (\xi_f/r)$, with $\chi \in (0, \pi)$, so $\chi$ is a monotonically decreasing function of the working frequency. Then combined with Eq. (C2) and (C6), we can conclude that $\psi_2$ is a monotonically increasing function of $\varphi$ while a monotonically decreasing function of the working frequency. Combined with the dependence of $\phi$ on $\psi_2$, we find $\phi$ is a monotonically decreasing function of the polar angle $\varphi$ and a monotonically increasing function of the working frequency. Combine Eq. (C6) and (C8), we know that when $\varphi + \chi$ goes from $-\pi$ to $\pi$, $\psi_2$ varies from $-\pi/2$ to $\pi/2$. Meanwhile, Eq. (C14) tells that when $\psi_1/2 + \psi_2 + k_a x - \pi/4$ varies from $-\pi/2$ to $\pi/2$, the reflection phase goes from $\pi$ to $-\pi$. So we can conclude that when the polar angle $\varphi$ circles the Weyl point once, the reflection phase goes continuously from $\pi$ to $-\pi$ once. This explains the vortex structure of the reflection phase in the parameter space and the charge of the vortex is the same as the charge of the Weyl point.

Also changing the truncation position or the working frequency just shift the reflection phases and do not change the conclusion above. As an example, we change the truncating position and now the PCs are truncated at the starting plane of the first layer. The corresponding reflection phases in the $p$-$q$ space are shown in Fig. 9(a), where the parameters used are the same as those in Fig. 3(a) and the working frequency is also fixed at the frequency of the Weyl point. Compared with Fig. 3(a), the reflection phase in Fig. 9(a) changes locally but the charge of the vortex is still preserved. If the truncating position of the PC is inside other layers, we can obtain the same conclusion following a similar proof as the one given above.

We now analyze Eq. (C14) in more detail. When $\xi_f = -r$, we have $\chi = \pi$, which is independent of the value of $r$, so the reflection phase maintains a constant for a fixed $\varphi$; while for $\xi_f = r$, we can also get the same result, but with $\chi = 0$ instead. Now we consider the dispersion of the interface states between PCs and a mirror with reflection phase of $0$ in the parameter space. As shown in Fig. 9(b), the interface states form a surface which rotates 180° around the Weyl point as frequency increasing. Such feature is discussed before [55], however as effective Hamiltonians are used in Fig. 9(b), the intersection between the interface states and the bulk states is a straight line here.

In the above derivation, we use the transfer matrix method which is specialized to our systems. However, the existence of the reflection phase vortex pinning at the Weyl point is not specialized to our systems. To see this point, let us focus on the band gap region. In general, the surface property of a truncated bulk is characterized by the impedance matrix [56]. However, if the surface property is dominated by one plane wave component and then the surface property can be approximately described by a scalar surface impedance or equivalently the reflection phase [57]. Let us then assume that inside the boundary region between a system with Weyl points and a reflecting substrate, there exist arc-like surface states originating from one Weyl point. Note that these surface states should always exist regardless of the property of the reflecting substrate, hence the reflection phase should winding in the periodic $[-\pi\ \pi)$ when circling around the Weyl point. This indicates that the reflection phase vortex pining at the Weyl point is a general feature in a system possessing Weyl points.

**References**

[1] X. Wan, A. M. Turner, A. Vishwanath, and S. Y. Savrasov, *Topological semimetal and Fermi-arc surface states in the electronic structure of pyrochlore iridates*, Phys. Rev. B **83**, 205101 (2011).

[2] P. Hosur and X. Qi, *Recent developments in transport phenomena in Weyl semimetals*, Comp. Rend. Phys. **14**, 857 (2013).

[3] B. Q. Lv *et al.*, *Experimental Discovery of Weyl Semimetal TaAs*, Phys. Rev. X **5**, 031013 (2015).

[4] S.-Y. Xu *et al.*, *Discovery of a Weyl fermion semimetal and topological Fermi arcs*, Science **349**, 613 (2015).

[5] A. A. Soluyanov, D. Gresch, Z. Wang, Q. Wu, M. Troyer, X. Dai, and B. A. Bernevig, *Type-II Weyl semimetals*, Nature **527**, 495 (2015).

[6] K. Deng *et al.*, *Experimental observation of topological Fermi arcs in type-II Weyl semimetal MoTe2*, Nat. Phys. **12**, 1105 (2016).

[7] T.-R. Chang *et al.*, *Prediction of an arc-tunable Weyl Fermion metallic state in MoxW1-xTe2*, Nat. Commun. **7**, 10639 (2016).

[8] S.-Y. Xu *et al.*, *Discovery of a Weyl fermion state with Fermi arcs in niobium arsenide*, Nat. Phys. **11**, 748 (2015).

[9] L. Lu, L. Fu, J. D. Joannopoulos, and M. Soljacic, *Weyl points and line nodes in gyroid photonic crystals*, Nat. Photon **7**, 294 (2013).

[10] B.-A. Jorge, L. Ling, F. Liang, B. Hrvoje, and S. Marin, *Weyl points in photonic-crystal superlattices*, 2D Mater. **2**, 034013 (2015).

[11] M. Xiao, Q. Lin, and S. Fan, *Hyperbolic Weyl Point in Reciprocal Chiral Metamaterials*, Phys. Rev. Lett. **117**, 057401 (2016).

[12] W.-J. Chen, M. Xiao, and C. T. Chan, *Photonic crystals possessing multiple Weyl points and the experimental observation of robust surface states*, Nat. Commun. **7**, 13038 (2016).

[13] M.-L. Chang, M. Xiao, W.-J. Chen, and C. T. Chan, *Multiple Weyl points and the sign change of their topological charges in woodpile photonic crystals*, Phys. Rev. B **95**, 125136 (2017).

[14] J. Noh, S. Huang, D. Leykam, Y. D. Chong, K. P. Chen, and M. C. Rechtsman, *Experimental observation of optical Weyl points and Fermi arc-like surface states*, Nat. Phys **13**, 611 (2017).

[15] M. Xiao, W. J. Chen, W. Y. He, and C. T. Chan, *Synthetic gauge flux and Weyl points in acoustic systems*, Nat. Phys. **11**, 920 (2015).

[16] Z. Yang and B. Zhang, *Acoustic Type-II Weyl Nodes from Stacking Dimerized Chains*, Phys. Rev. Lett. **117**, 224301 (2016).

[17] W. Gao, B. Yang, M. Lawrence, F. Fang, B. Beri, and S. Zhang, *Photonic Weyl degeneracies in magnetized plasma*, Nat. Commun. **7**, 12435 (2016).

[18] P. Roushan *et al.*, *Observation of topological transitions in interacting quantum circuits*, Nature **515**, 241 (2014).

[19] M. D. Schroer, M. H. Kolodrubetz, W. F. Kindel, M. Sandberg, J. Gao, M. R. Vissers, D. P. Pappas, A. Polkovnikov, and K. W. Lehnert, *Measuring a Topological Transition in an Artificial Spin-$1/2$ System*, Phys. Rev. Lett. **113**, 050402 (2014).

[20] R.-P. Riwar, M. Houzet, J. S. Meyer, and Y. V. Nazarov, *Multi-terminal Josephson junctions as topological matter*, Nat. Commun. **7**, 11167 (2016).

[21] O. Boada, A. Celi, J. I. Latorre, and M. Lewenstein, *Quantum Simulation of an Extra Dimension*, Phys. Rev. Lett. **108**, 133001 (2012).

[22] L. Yuan, Y. Shi, and S. Fan, *Photonic gauge potential in a system with a synthetic frequency dimension*, Opt. Lett. **41**, 741 (2016).

[23] B. Lian and S.-C. Zhang, *Five-dimensional generalization of the topological Weyl semimetal*, Phys. Rev. B **94**, 041105 (2016).

[24] M. Feng, X. Zheng-Yuan, Z. Dan-Wei, T. Lin, L. Chaohong, and Z. Shi-Liang, *Witnessing topological Weyl semimetal phase in a minimal circuit-QED lattice*, Quantum. Sci. Technol. **1**, 015006 (2016).

[25] Q. Lin, M. Xiao, L. Yuan, and S. Fan, *Photonic Weyl Point in a Two-Dimensional Resonator Lattice with a Synthetic Frequency Dimension*, Nat. Commun. **7**, 13731 (2016).

[26] L. Lu, Z. Wang, D. Ye, L. Ran, L. Fu, J. D. Joannopoulos, and M. Soljačić, *Experimental observation of Weyl points*, Science **349**, 622 (2015).

[27] L. Lu and Z. Wang, *Topological one-way fiber of second Chern number* in *arXiv: 1611.01998* (2016).

[28] Z. Fang *et al.*, *The Anomalous Hall Effect and Magnetic Monopoles in Momentum Space*, Science **302**, 92 (2003).

[29] N. Yu, P. Genevet, M. A. Kats, F. Aieta, J.-P. Tetienne, F. Capasso, and Z. Gaburro, *Light Propagation with Phase Discontinuities: Generalized Laws of Reflection and Refraction*, Science **334**, 333 (2011).

[30] M. Xiao, Z. Q. Zhang, and C. T. Chan, *Surface Impedance and Bulk Band Geometric Phases in One-Dimensional Systems*, Phys. Rev. X **4**, 021017 (2014).

[31] Q. Wang, M. Xiao, H. Liu, S. Zhu, and C. T. Chan, *Measurement of the Zak phase of photonic bands through the interface states of a metasurface/photonic crystal*, Phys. Rev. B **93**, 041415 (2016).

[32] M. S. Rudner, N. H. Lindner, E. Berg, and M. Levin, *Anomalous Edge States and the Bulk-Edge Correspondence for Periodically Driven Two-Dimensional Systems*, Phys. Rev. X **3**, 031005 (2013).

[33] G. Lheureux, S. Azzini, C. Symonds, P. Senellart, A. Lemaitre, C. Sauvan, J. P. Hugonin, J. J. Greffet, and J. Bellessa, *Polarization-Controlled Confined Tamm Plasmon Lasers*, ACS Photonics **2**, 842 (2015).

[34] C. Symonds, G. Lheureux, J. P. Hugonin, J. J. Greffet, J. Laverdant, G. Brucoli, A. Lemaitre, P. Senellart, and J. Bellessa, *Confined Tamm Plasmon Lasers*, Nano Lett. **13**, 3179 (2013).

[35] C. Symonds, A. Lemaitre, P. Senellart, M. H. Jomaa, S. A. Guebrou, E. Homeyer, G. Brucoli, and J. Bellessa, *Lasing in a hybrid GaAs/silver Tamm structure*, Appl. Phys. Lett. **100**, 121122 (2012).

[36] O. Gazzano, S. M. de Vasconcellos, K. Gauthron, C. Symonds, P. Voisin, J. Bellessa, A. Lemaitre, and P. Senellart, *Single photon source using confined Tamm plasmon modes*, Appl. Phys. Lett. **100**, 232111 (2012).

[37] O. Gazzano, S. M. de Vasconcellos, K. Gauthron, C. Symonds, J. Bloch, P. Voisin,

J. Bellessa, A. Lemaître, and P. Senellart, *Evidence for Confined Tamm Plasmon Modes under Metallic Microdisks and Application to the Control of Spontaneous Optical Emission*, Phys. Rev. Lett. **107**, 247402 (2011).

[38] Z. Y. Yang, S. Ishii, T. Yokoyama, T. D. Dao, M. G. Sun, T. Nagao, and K. P. Chen, *Tamm plasmon selective thermal emitters*, Opt. Lett. **41**, 4453 (2016).

[39] H. Lu, X. T. Gan, B. H. Jia, D. Mao, and J. L. Zhao, *Tunable high-efficiency light absorption of monolayer graphene via Tamm plasmon polaritons*, Opt. Lett. **41**, 4743 (2016).

[40] V. Villafane, A. E. Bruchhausen, B. Jusserand, P. Senellart, A. Lemaitre, and A. Fainstein, *Confinement of gigahertz sound and light in Tamm plasmon resonators*, Phys. Rev. B **92**, 165308 (2015).

[41] B. Auguie, M. C. Fuertes, P. C. Angelome, N. L. Abdala, G. Illia, and A. Fainstein, *Tamm Plasmon Resonance in Mesoporous Multilayers: Toward a Sensing Application*, ACS Photonics **1**, 775 (2014).

[42] Y. K. Chen *et al.*, *Back focal plane imaging of Tamm plasmons and their coupled emission*, Laser Photonics Rev. **8**, 933 (2014).

[43] G. Bräunlich, G. M. Graf, and G. Ortelli, *Equivalence of Topological and Scattering Approaches to Quantum Pumping*, Commun. Math. Phys. **295**, 243 (2010).

[44] I. C. Fulga, F. Hassler, and A. R. Akhmerov, *Scattering theory of topological insulators and superconductors*, Phys. Rev. B **85**, 165409 (2012).

[45] H. Wang, L. Zhou, and Y. D. Chong, *Floquet Weyl phases in a three-dimensional network model*, Phys. Rev. B **93**, 144114 (2016).

[46] Z. Yan and Z. Wang, *Tunable Weyl Points in Periodically Driven Nodal Line Semimetals*, Phys. Rev. Lett. **117**, 087402 (2016).

[47] H. B. Nielsen and M. Ninomiya, *Absence of neutrinos on a lattice*, Nucl. Phys. B **193**, 173 (1981).

[48] N. D. Mermin, *The topological theory of defects in ordered media*, Rev. Mod. Phys. **51**, 591 (1979).

[49] C. Fang, Y. Chen, H.-Y. Kee, and L. Fu, *Topological nodal line semimetals with and without spin-orbital coupling*, Phys. Rev. B **92**, 081201 (2015).

[50] K. F. MacDonald, Z. L. Samson, M. I. Stockman, and N. I. Zheludev, *Ultrafast active plasmonics*, Nat. Photon **3**, 55 (2009).

[51] D. A. Fuhrmann, S. M. Thon, H. Kim, D. Bouwmeester, P. M. Petroff, A. Wixforth, and H. J. Krenner, *Dynamic modulation of photonic crystal nanocavities using gigahertz acoustic phonons*, Nat. Photon **5**, 605 (2011).

[52] S. Forstner, S. Prams, J. Knittel, E. D. van Ooijen, J. D. Swaim, G. I. Harris, A. Szorkovszky, W. P. Bowen, and H. Rubinsztein-Dunlop, *Cavity Optomechanical Magnetometer*, Phys. Rev. Lett. **108**, 120801 (2012).

[53] C. Sheng, H. Liu, S. Zhu, and D. A. Genov, *Active control of electromagnetic radiation through an enhanced thermo-optic effect*, Sci. Rep. **5**, 8835 (2015).

[54] C. Fang, M. J. Gilbert, X. Dai, and B. A. Bernevig, *Multi-Weyl Topological Semimetals Stabilized by Point Group Symmetry*, Phys. Rev. Lett. **108**, 266802 (2012).

[55] C. Fang, L. Lu, J. Liu, and L. Fu, *Topological semimetals with helicoid surface states*, Nat. Phys. **12**, 936 (2016).

[56] F. J. Lawrence, L. C. Botten, K. B. Dossou, R. C. McPhedran, and C. M. de Sterke, *Photonic-crystal surface modes found from impedances*, Phys. Rev. A **82**, 053840 (2010).

[57] X. Huang, M. Xiao, Z.-Q. Zhang, and C. T. Chan, *Sufficient condition for the existence of interface states in some two-dimensional photonic crystals*, Phys. Rev. B **90**, 075423 (2014).

## Acknowledgements

H.L. gratefully acknowledges the support of the National Natural Science Foundation of China (No's 11321063, 61425018 and 11374151). C.T.C. and M. X. gratefully acknowledge the support of the Research Grants Council, University Grants Committee, Hong Kong (AoE/P-02/12).

## Author contributions

Q.W. and M.X. proposed and designed the system. Q.W., H.L. and S.Z. carried out the experiments. Q.W., M.X., C.T.C., H.L. contributed to the experimental characterization and interpretation, and developed the theory. Q.W. M.X. and C.T.C. co-wrote the manuscript. All the authors were involved in the discussions.

## Competing financial interests

**The authors declare no competing financial interests.**

**Figures**

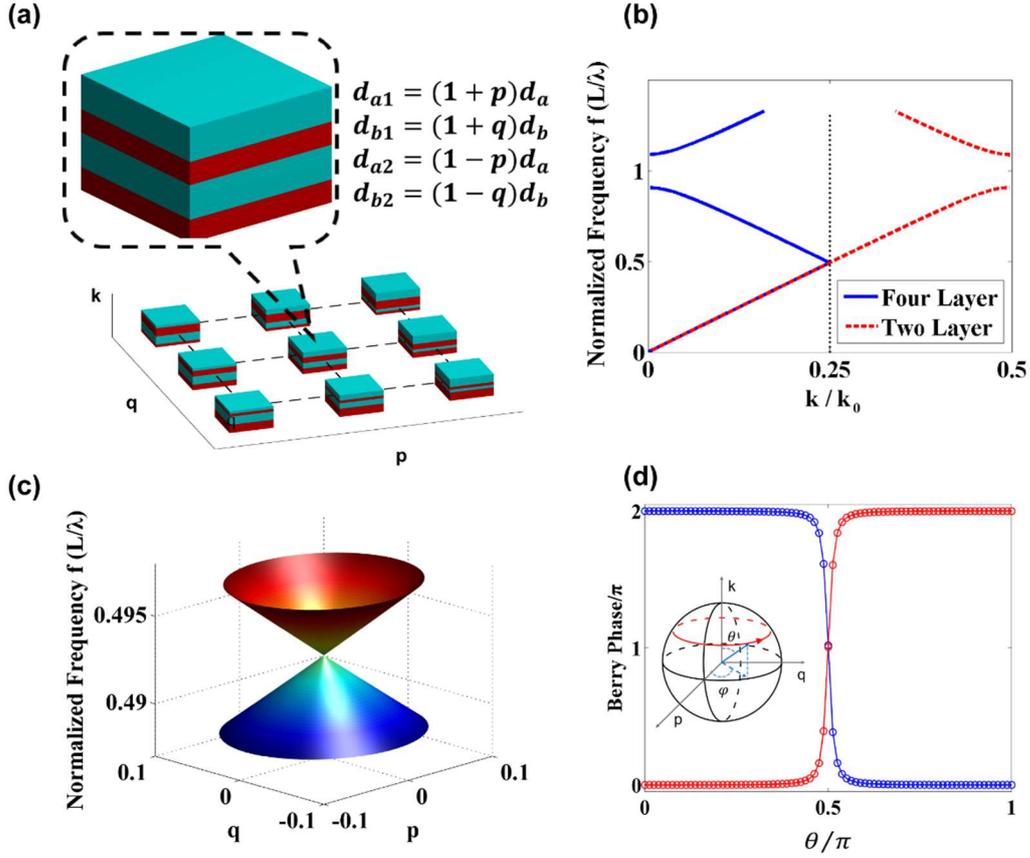

FIG. 1. Realization of Weyl points in a parameter space. (a) Photonic crystals (PCs) with different $p, q$ values. The $p, q$ form a parameter space. The inset shows one unit cell of the PC, where the first and the third layers are made of $H_fO_2$ (blue), and the second and the fourth layers are made of $S_iO_2$ (red). The thickness of each layer depends on its position in the $p$-$q$ parameter space. (b) The band dispersion of PCs with different unit cells (red line for a PC with only two layers and blue line for a PC the unit cell of which consists of two unit cells used for the red line). Here $d_a = 97 nm$ and $d_b = 72 nm$. (c) The dispersion of PCs in the $p$-$q$ space with $k = 0.5 k_0$. Here two bands form a conical intersection. (b) and (c) together show that the band dispersions are linear in all directions around the degeneracy point, indicating that it is an analogue of the Weyl point. (d) Berry phases defined on the spherical surface with a fixed $\theta$ (see the inset), where blue and red represent

Berry phases on the lower and the upper band, respectively. The radius of the sphere is 0.001.

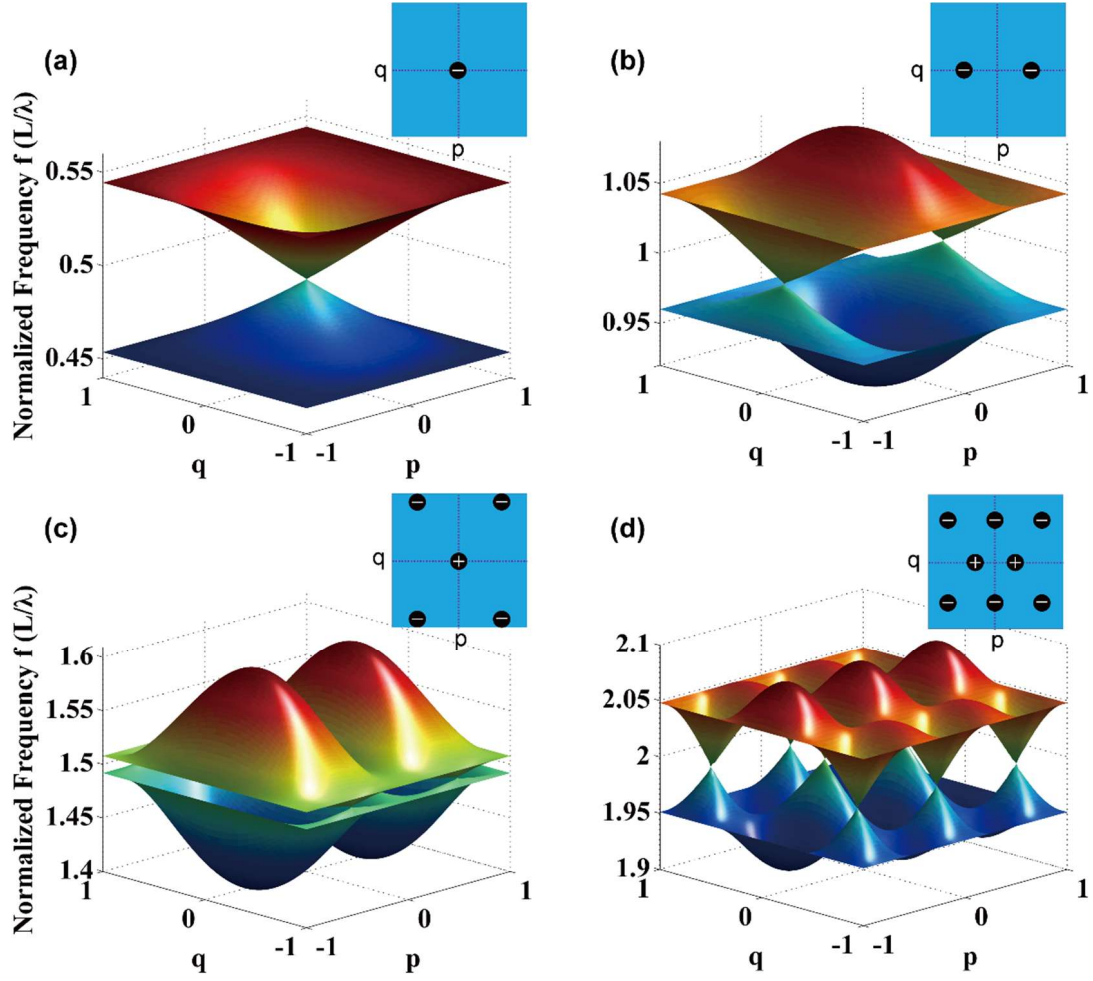

FIG. 2. The band dispersions in the parameter space at different k points. (a) for bands 1 and 2 at $k = 0.5k_0$. (b) for bands 2 and 3 at $k = 0$. (c) for bands 3 and 4 at $k = 0.5k_0$. (d) for bands 4 and 5 at $k = 0$. The conical intersections in (a)-(d) all correspond to Weyl points with positions and charges ("+" for charge +1, "-" for charge -1) marked in the insets, in which the dashed lines denote $p = 0$ or $q = 0$. Here we use the same parameters as those in Fig. 1.

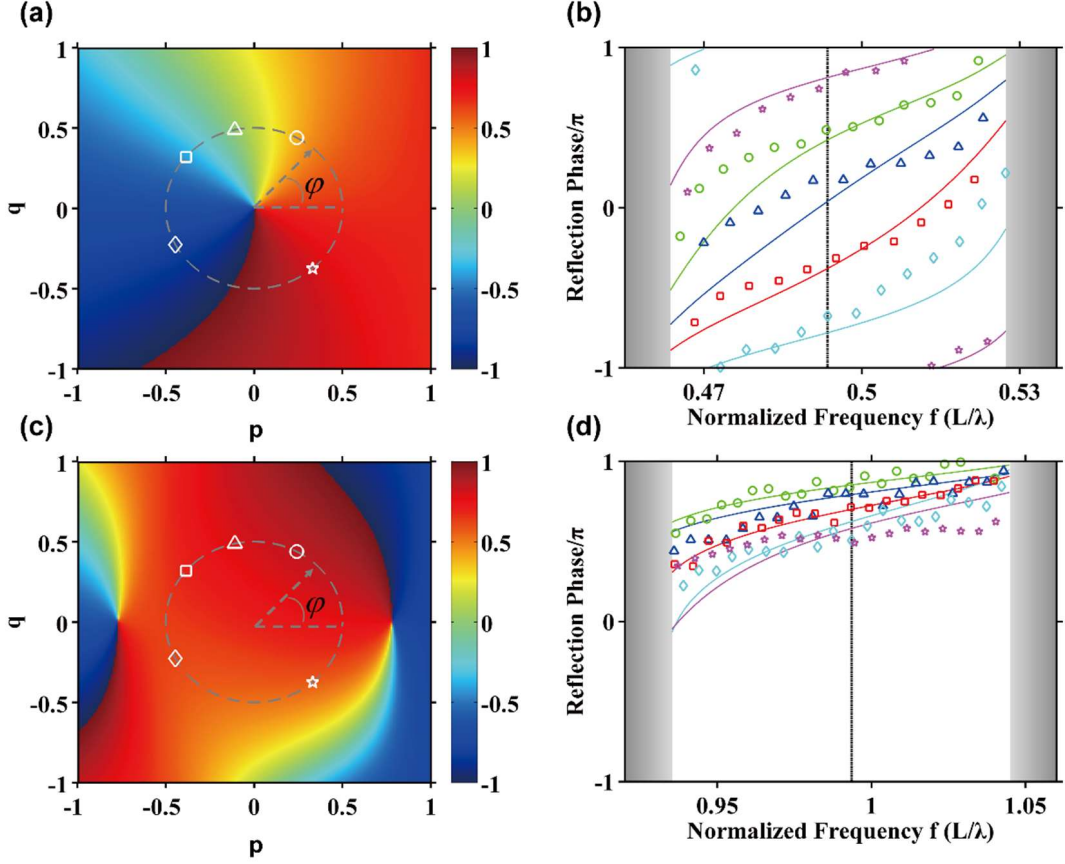

FIG. 3. The reflection phase in the $p\text{-}q$ space. (a) and (c) the reflection phase in the $p\text{-}q$ space at the frequency of the Weyl point in Fig. 2(a) and Fig. 2(b). The reflection phase shows a vortex structure with charge -1 around Weyl points. The white circle, triangle, square, diamond and pentagram mark the values of $(p,q)$ of the five samples, which are respectively, $(p,q) = (0.24, 0.44)$, $(-0.11, 0.49)$, $(-0.38, 0.31)$, $(-0.45, 0.22)$ and $(0.33, -0.37)$. These five samples have configurations located on a circle (dashed gray line) in the parameter space which encloses a Weyl point in (a) but does not enclose any Weyl point in (c). (b) and (d) reflection phases in the band gaps of the five samples in (a) and (c). The markers show the experimental results whose shape mark the position in the $p\text{-}q$ space in (a) and (c), and colored lines represent the corresponding reflection phases from numerical simulations. The black dashed lines mark the frequencies of the Weyl points in (a) and (c), and the bulk band regions of the PC with the narrowest

band gap are shaded in gray.

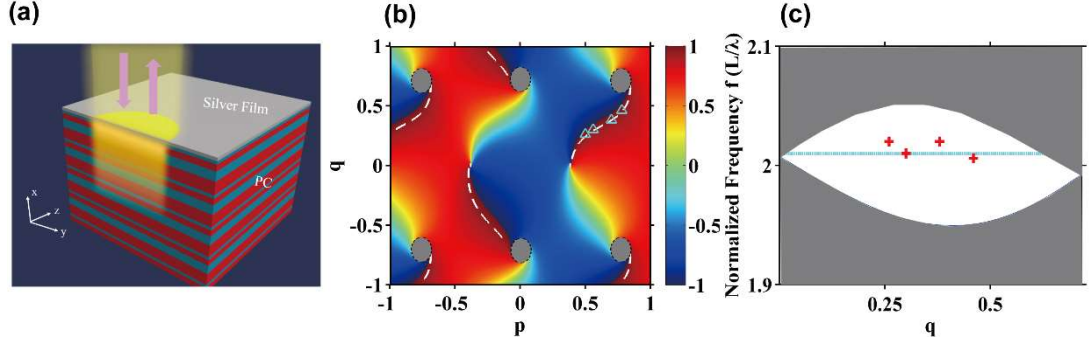

FIG. 4. The 'Fermi arc' in the parameter space. (a) A sketch shows the structure used in measuring the interface states. Here the silver slab represents the silver film, blue represents $H_fO_2$ and red represents $S_iO_2$. Wave incidents from the silver film side and the interface states are localized in the interface region between the silver slab and the semi-infinite PC. The amplitude of the wave is shown schematically in yellow. (b) The reflection phase of the PCs at the frequency of Weyl points with charge +1 with its dispersion shown in Fig. 2(d). The bulk band regions at the working frequency are shaded in gray. The white dashed lines show the trajectories of the interface states of a system consisting of a semi-infinite PC coated with a silver film, where the semi-infinite PC is truncated at the center of the first layer. These interface states trajectories are analogues of Fermi arc states. The triangles mark the $p$ and $q$ values of the four samples in the experiment. (c) The cyan line indicates the working frequency used in (b), and the red crosses label the experimental results. The projection of the bulk band as a function of $q$ is shaded in gray. The number of unit cells is 10, and $d_a = 0.323um$ and $d_b = 0.240um$ for the four PCs in (c). The thickness of layers in these four PCs are given by $(p,q) = (0.50, 0.26)$, $(0.56, 0.30)$, $(0.70, 0.38)$ and $(0.78, 0.36)$ respectively.

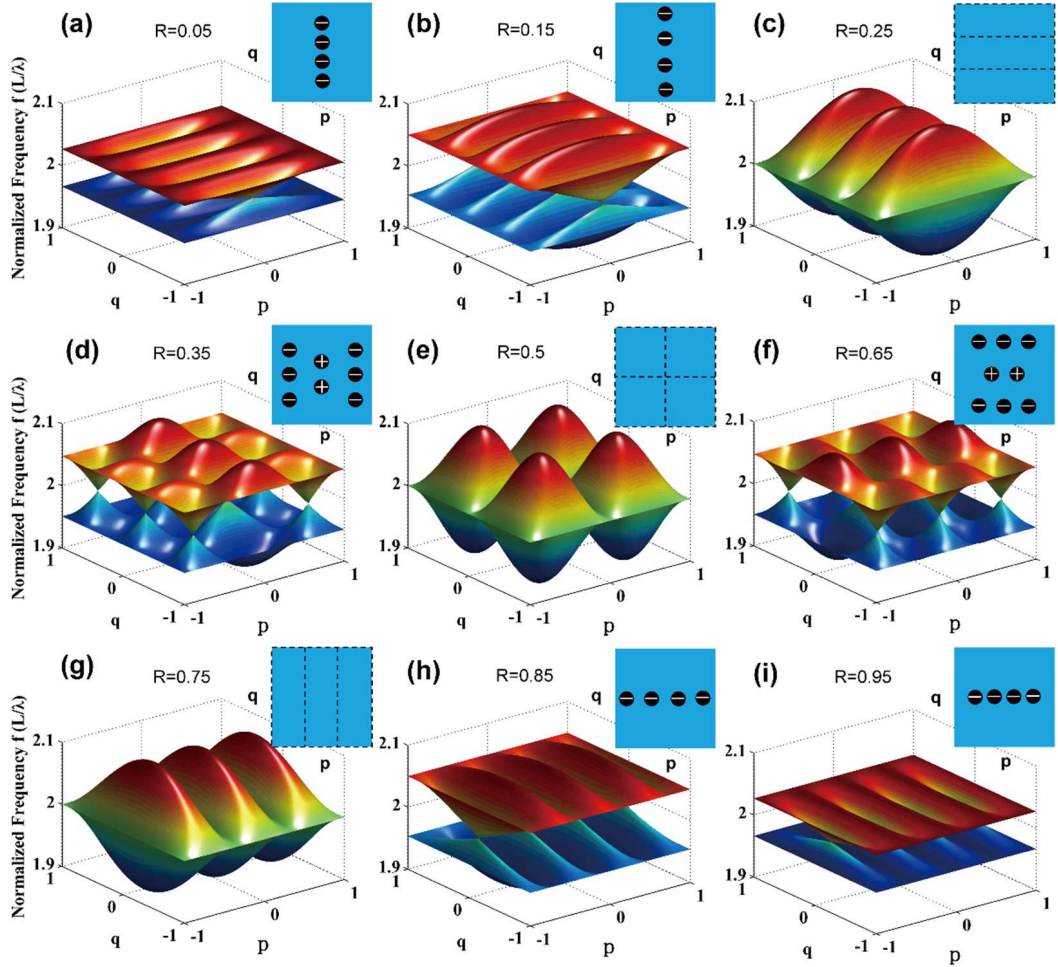

FIG. 5. Topological transitions occur as we change the parameter R defined in the text. (a)-(i), the band structures of band 4 and band 5 with different values of R at $k=0$. The values of R are S0.05, 0.15, 0.25, 0.35, 0.5, 0.65, 0.75, 0.85, and 0.95 for (a)-(i), respectively. The insets show the positions of the Weyl points ("+" for positive charge and "-" for negative charge) and "nodal lines" (black dashed lines).

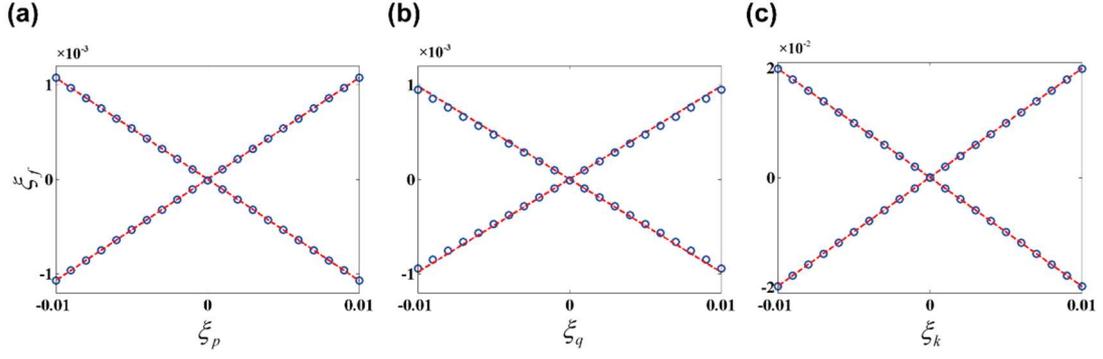

FIG. 6. The dispersion near the Weyl points. (a), (b), (c), show the comparison between the numerical result (blue circles) and the effective Hamiltonian in Eq. (A16) (red dashed lines) in three directions. The parameters of the PC are given by $d_a = 97nm$, $d_b = 72nm$, and the Weyl point considered is between the first and the second bands.

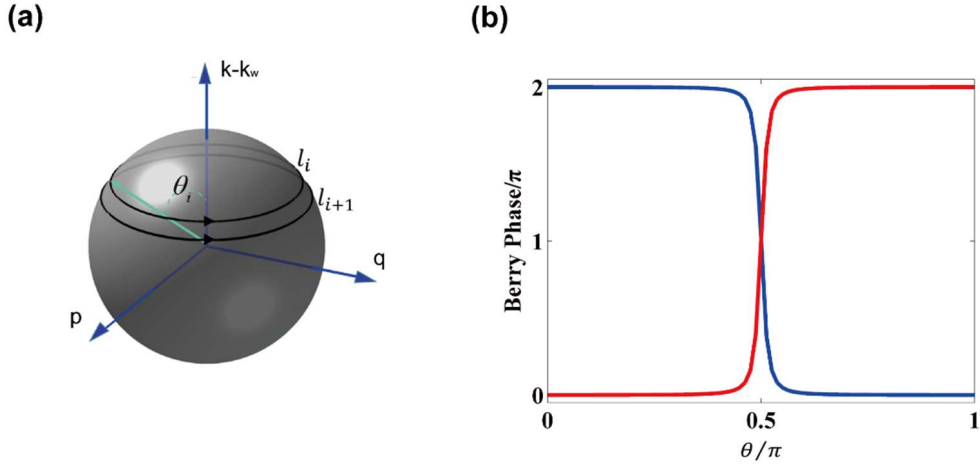

FIG.7. The method to calculate the charge of Weyl point numerically. (a), a schematic illustration of the integration paths used to calculate topological charges of Weyl points. (b), the Berry phases defined on the spherical surface with fixed $\theta$, where blue and red represent Berry phases on the lower band and the upper band, respectively. The parameters of the PC are given by $d_a = 97nm$, $d_b = 72nm$, and the Weyl point considered is between the first and the second band. The radius of the sphere is 0.001.

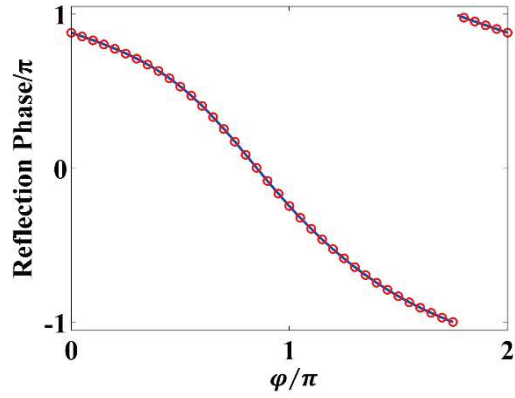

FIG. 8. The reflection phase as a function of the polar angle $\varphi$ on a circle with radius 0.01 in the $p\text{-}q$ space around the Weyl point. Blue solid line and red circles represent the reflection phases calculated through the transfer matrix method and Eq. (C14), respectively. Here we consider the same Weyl point as in Fig. 6. The parameters of the PC is given by $d_a = 97nm$, $d_b = 72nm$, and the Weyl point considered is between the first and the second bands and located at $(p,q)=(0,0)$. The PC is truncated at the starting plane of the first layer, and the working frequency is also fixed at the frequency of the Weyl point, which is $f_w = 247.6THz$.

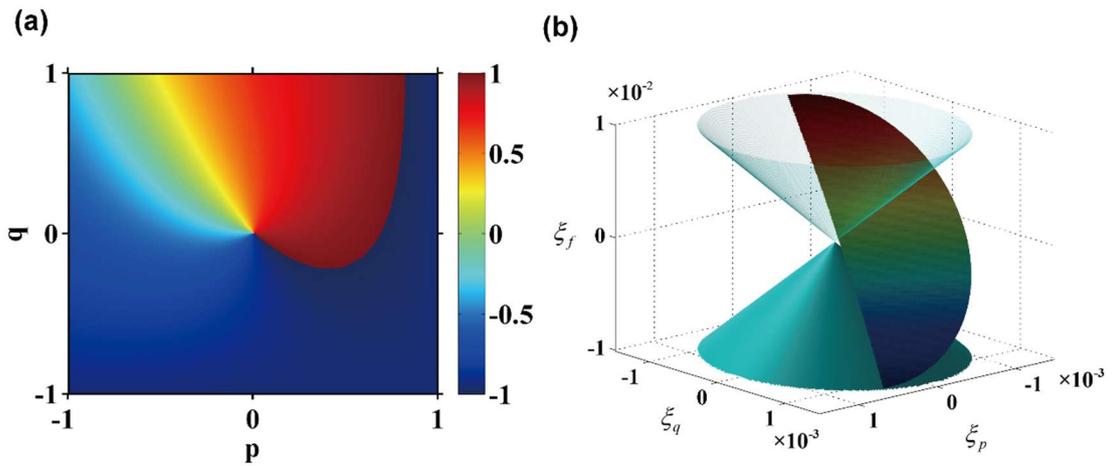

FIG. 9. (a), The reflection phase in the $p\text{-}q$ space for the PCs with $d_a = 97nm$ and $d_b = 72nm$. Now the PCs are truncated at the starting plane of the first layer. The working frequency is also fixed at the frequency of the Weyl point, which is $f_w = 247.6\text{THz}$. Compared with Fig. 3(a), the reflection phase in the $p$-$q$ space changes locally, but the charge of the vortex located at the Weyl point keeps the same. (b), The dispersion of interface states near a Weyl point, where the cyan cone represents the bulk state and the colored sheet represents the interface states. Color code of the interface states represents different frequencies. In this calculation, the photonic crystals are truncated at the starting plane of the first layer.

# Supplementary Material for

"

# Optical interface states protected by synthetic Weyl points

"

## Section I: Reflection phase measurement

To measure the reflection phases, we adopt a Fabry–Perot (FP) interference method[1,2]. The FP cavity consists of an air cavity sandwiched by two mirrors. The inset in Fig. S2(a) and Fig. S2(a) show respectively, the experimental setups to measure the reflection phases of the silver film and the PCs. As the thickness of the glass slab is much larger than the wavelength, the interference coming from the upper glass boundary can be ignored. Use the two beams approximation, the condition of the FP resonance peak is given by:

$$\phi_1 + \phi_2 + 4\pi d \frac{f_m}{c} = 2m\pi, \qquad m \in \mathbb{Z} \tag{1}$$

where $\phi_1$ and $\phi_2$ represent the reflection phases of the mirrors above the air cavity and below the air cavity, respectively, $d$ is the thickness of the air gap, $f_m$ is frequency of m$^{th}$ resonance, and $c$ is the speed of light in vacuum. If $\phi_1$ and $\phi_2$ are independent of the working frequency, we can have:

$$d = \frac{c}{\Delta f}, \tag{2}$$

where $\Delta f$ is the frequency difference between two neighboring resonances peaks.

We start with the measurement of the reflection phase of the silver film to show how to obtain the reflection phase in more detail. The measured refection phase of the silver film is used to determine the position of the interface states in Fig. 4(a) (the white dashed lines). Here the working frequency is at around 303THz. We neglect the dispersion of the silver film as the frequency range of interest is narrow. The thickness of the two silver films in this measurement are estimated to be around 20nm. Fig. S1(a) shows the measured reflection spectrum. The distance between these two silver films is obtained through fitting with Eq. (2). Substitute into Eq. (1), we now obtained the reflection phase of the silver films as shown in Fig. S1(b), which is about $(-0.95\pm 0.0471)\pi$ around 303THz.

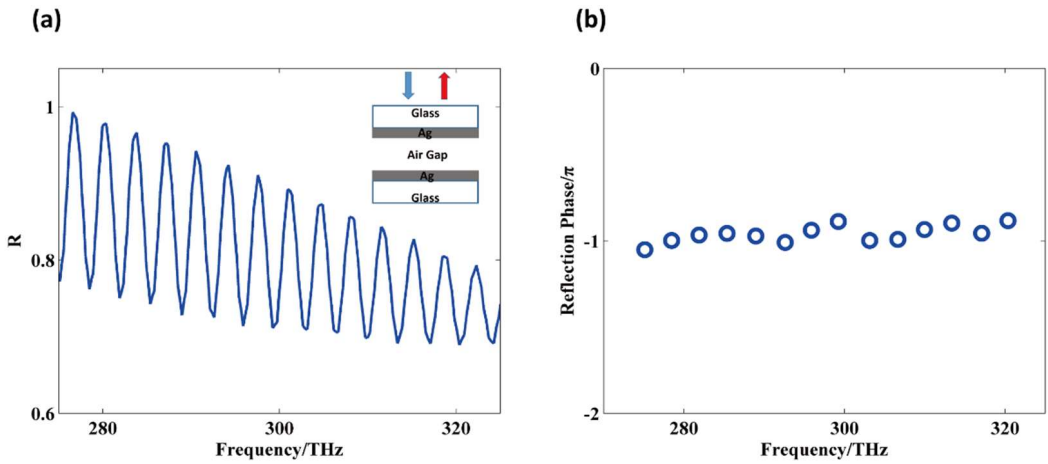

Figure S1 (a), the reflection spectrum of the Fabry–Perot (FP) cavity composed of two silver films, and the corresponding experimental setup is shown in the inset. The reflection spectrum is measured above the top glass. (b), the deduced reflection phase of silver film from the reflection spectrum in (a).

The measurement of the reflection phase of the PCs is similar to the above discussion. The setup is shown in Fig. S2(a). Half of the PCs are deposited with silver films with thickness $200nm\pm 10nm$. This silver film is thick enough to reflect all the incident waves with reflection phase $\pi$. The silver film above the air gap is designed to be 7nm such that the incident wave can partially pass through it. The first measurement is performed inside the region where the FP cavity is composed of the two silver films, through which we can obtain the distance between these two silver films and the reflection phase of the silver film above the air gap. The reflection phase of the silver films above the air gap are measured to be $(-0.41\pm 0.039)\pi$ inside the first band gap and $(-0.40 \pm 0.111)\pi$ inside the second band gap in Fig. 3. With the thickness of air gap and the reflection phase of the silver film above the air gap determined, the reflection phase of the PCs can now be obtained through a secondary measurement inside the region where the FP cavity consists of the silver film and the PCs.

Fig.S2(b) shows the SEM image of a PC designed using parameters: $d_a = 97 nm$, $d_b = 72 nm$, $p = -0.38$ and $q = -0.31$. The PC is truncated using the Focus Ion Beam and the angle between the PC and the observation view here is 52°. Here light gray strips are $H_fO_2$ while the dark strips are $S_iO_2$. Fig. S2(b) shows that each layer of the PC is uniform with some small local variation. We expect a small global shift between the thickness of the designed and the fabricated samples. We use $\alpha$ to denote this global shift, then

$$L_f = (1+\alpha) L_d, \qquad (3)$$

where $L_f$ and $L_d$ represent the optical length of the fabricated sample and the designed value. $\alpha$ introduces a shift of the band gap frequency range. Hence we can obtain the value of $\alpha$ by fitting the frequency of the band gap while keeping the value of $p$ and $q$. In Fig. S2(c), we show the measured reflection spectrum (red curve) and the numerical reflection spectrum with fitted thickness (blue curve) of a PC. The designed value of the thickness of each unit cell is $L_d = 597.0 nm$ while $L_f = 596.4 nm$, this gives $\alpha \approx -0.1\%$. The average value of $\alpha$ in our experiment is $0.72\%$ and the mean square difference is $1.3\%$. After subtracting the effect of this small global shift, we are now able to compare the reflection phases and the frequencies of the interface states with numerical simulation. As an example, we show reflection phases from experimental measurements (red diamonds) and from numerical simulation (blue curve) of one PC in Fig. S3(d). For this PC, the experimental measurement agrees well with the numerical simulation.

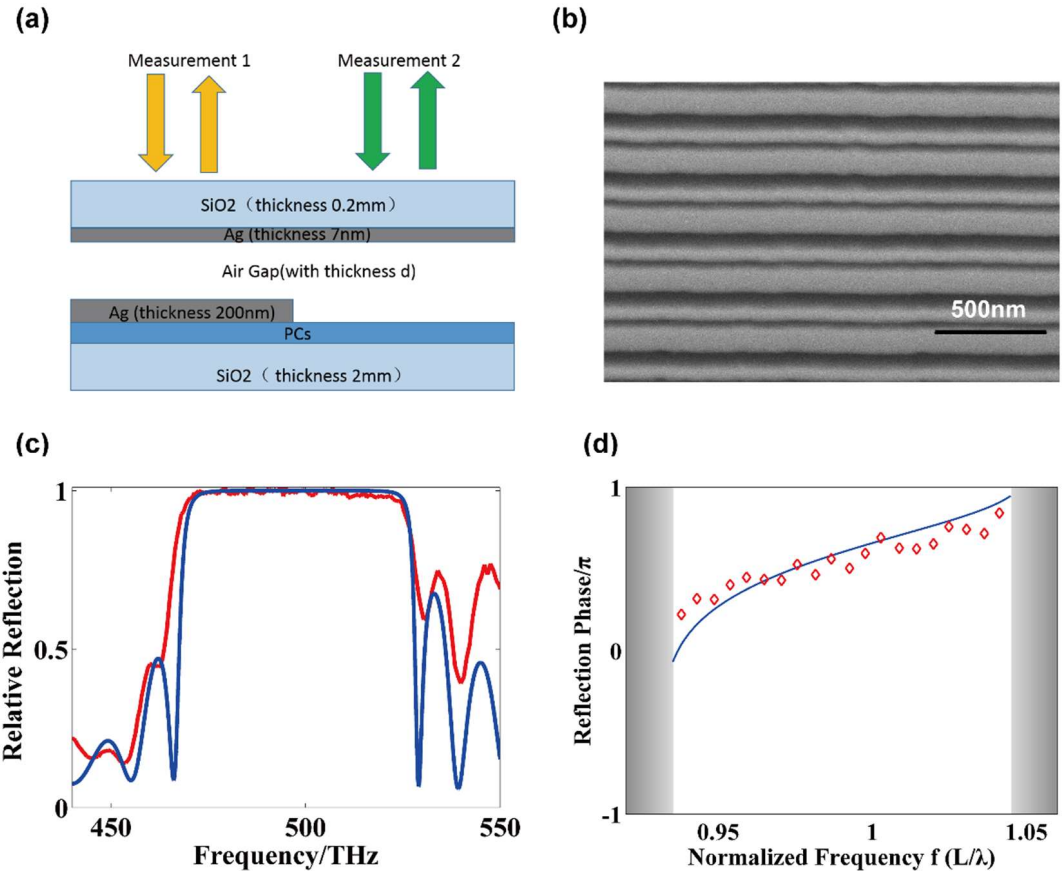

Figure S2 (a), the experimental setup to measure the reflection phase of PCs. (b), SEM image of a PC. (c), the red line shows the measured reflection spectrum of the same PC and the blue line represents the calculated reflection spectrum with $L_f = 596.4\text{nm}$. (d), the reflection phase of the same PC, where red diamonds represent experimental data, the blue line is from numerical simulation and gray strips represent pass band regions. The geometric parameters of the PC used in this figure is given by $d_a = 97nm$, $d_b = 72nm$, $p = -0.38$ and $q = -0.31$.

## Section II: Interface State Measurement

The interface states are at the boundary between the PCs and a silver films. In these measurements, we deposit 20nm-thick silver films on the top of the PCs. And the PCs are truncated at the half plane of the first layer. A sketch of the experimental setup is shown in Fig. S3(a), where the gray, blue and red layers represent silver, $H_fO_2$ and $S_iO_2$, respectively. The reflection spectrum is measured on the top of the silver layer. An example of the reflection spectrum is shown in Fig. S3(b) with the blue line. Here the gray vertical strips represent the bulk band regions. As the interface state enhances the absorption, the interface state here corresponds to the reflection dip inside the gap region as indicated by the red triangles in Fig. S3(b).

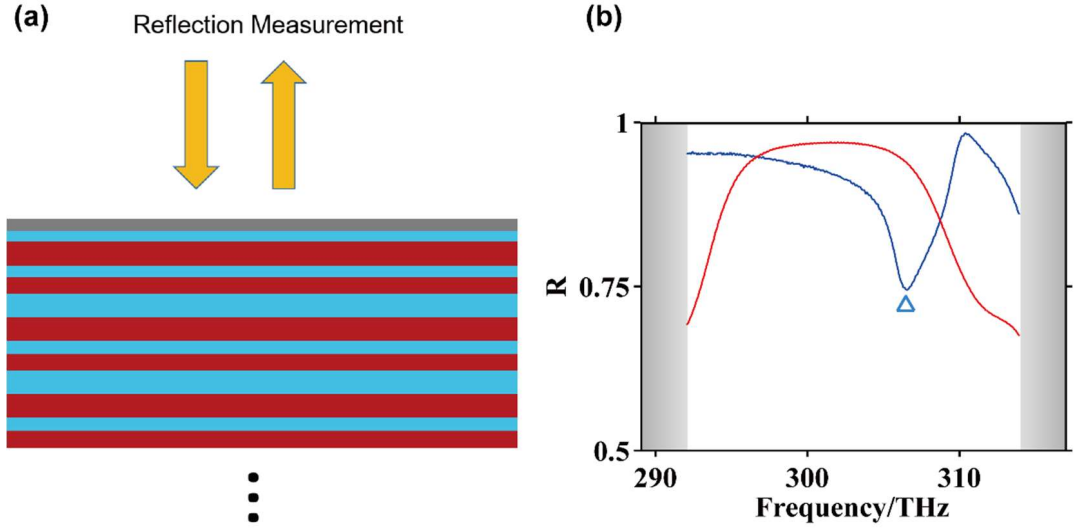

Figure S3 Measurement of the interface states. (a), The experimental setup to measure the frequencies of the interface states, where the gray, blue and red layers represent silver, $H_fO_2$ and $S_iO_2$, respectively. (b), The blue line represents the measured reflection spectrum. Here the parameters of the PC are given by $d_a = 0.323um$, $d_b = 0.240um$, $p = 0.56$ and $q = 0.34$, and the thickness of silver film is 20nm. The gray vertical strips represent the bulk band regions, and the dip (indicated by the red triangle) inside the band gap corresponds to the frequency of the interface state.

# Section III: The robustness of the 'Fermi arc like interface states'

Although we have only shown the 'Fermi arc like interface states' between PCs and silver films at the frequency of the Weyl points with charge +1 in the text, the trajectory of interface states actually extend to all the frequencies in the gap region. In Fig. S4(a) and (b), we show respectively, the reflection phases in the *p-q* space with the frequency of the Weyl point with charge -1 and a frequency above those of all the Weyl points. The vortex structures can be seen in both frequencies and they guarantee the existence of interface states. The interface states form a helicoid surface and such behavior has also been observed in Weyl semimetals[3].

In the main text, we have proved that the interface states always exist independent of the properties of the reflecting substrate. However, the trajectory of the interface states must connect one Weyl point to another Weyl point of an opposite charge, but which Weyl point it connects to actually depends on the property of the reflecting substrate (i.e. how the boundary is set up). As an example, we consider two reflecting substrates with reflection phases of 0 and $\pi$, respectively. Black lines in Fig. S4(a) represent the interface states when the reflection phase of the reflecting substrate is 0; while white lines in Fig. S4(b) represent the interface states when the reflection phase of the mirror is $\pi$. Indeed, the interface states connect different Weyl points as the reflection phase of the reflecting substrate changes.

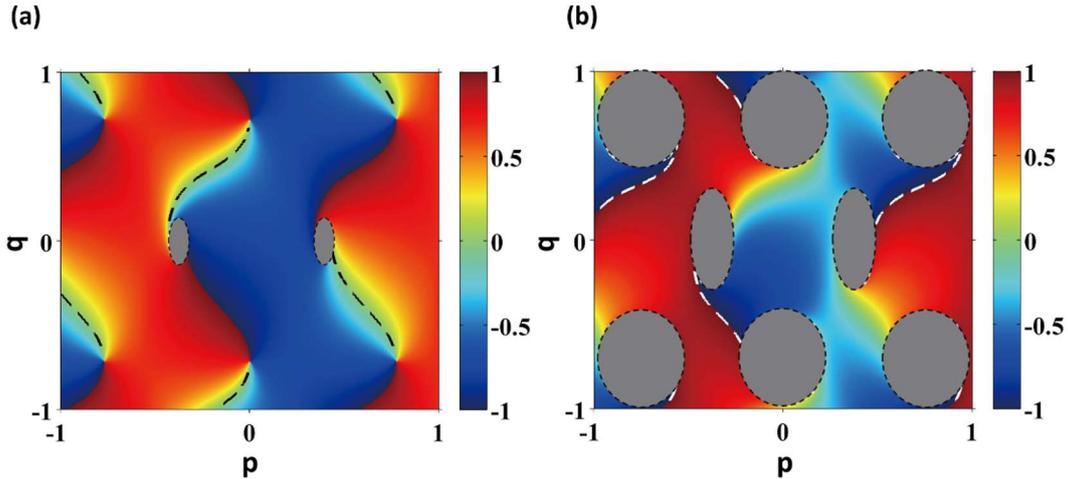

Figure S4 The color code in (a) and (b) represent the reflection phase (divided by $\pi$) in the *p-q* space, where the working frequencies are set to be the frequency of the Weyl point with charge -1 in (a) and a frequency above those of all the Weyl points in (b). The gray regions encircled by the dashed black lines represent the bulk band regions at the working frequency. The black lines in (a) and the white lines in (b) represent the trajectories of interface states formed by the PC and a mirror, where the reflection phase is 0 in (a) and $\pi$ in (b). The parameters of the PC are given by $d_a = 0.323um$, $d_b = 0.240um$ and the working frequencies are $f_0 = 300.3 \text{THz}$ and $f_0 = 307 \text{THz}$ in (a) and (b), respectively.

## Section IV: Absorption of interface states

The optical interface states are widely used in various systems [4], as its strong confinement of light will enhance the interaction of light and matters. The synthetic Weyl points proposed here give a flexible way to construct optical interface states between PCs and arbitrary reflecting substrates. As an example, we consider a system consisting of a PC and a 15nm-thick silver film. This system exhibits a interface states between the PC and the silver film. The field intensity of this system is shown in Fig. S5(a) as a function of position and the frequency of the incident wave. The optical wave is incident from the left side and propagate along the positive X direction, and the white dashed line indicates the interface between the PC and the silver film. The parameters of the PC are $d_a = 97nm$, $d_b = 72nm$, $p = 0.09$ and $q = -0.51$. The PC has 80 unit cells and is truncated at the center of the first layer. The relative permittivity of silver at the work frequency is $\varepsilon_{Ag} = -71.972 + 5.586i$ (from Palik database [5]). We can see that the intensity is greatly enhanced at the frequency of the cavity mode (f=247.6 THz), which falls inside the band gap region whose edge frequencies are labeled by the black dashed lines in Fig. S5(a). As the field intensities are greatly enhanced at the cavity mode resonances, interface states behavior as absorption peaks in the absorption spectra. Now we consider the same PCs-Ag film configurations but with different *p* and *q* values, and the absorption in the *p-q* space is shown in Fig. S5(b). Here the working frequency is fixed at the frequency of the synthetic Weyl point. The trajectory of the interface states at this fixed frequency is also shown in Fig. S5(b) with the black triangles. We can see the maximum of absorption can reach 95%, and the trajectory of the absorption peaks coincide with that of the interface states. There is a small derivation near the synthetic Weyl point between those two trajectories due to the finite size of the PCs used in the simulations (The gaps vanishes at the Weyl point). The $p$ and $q$ values of the PC used in Fig. S5(a) are also marked by the white triangle in Fig. S5(b) for reference.

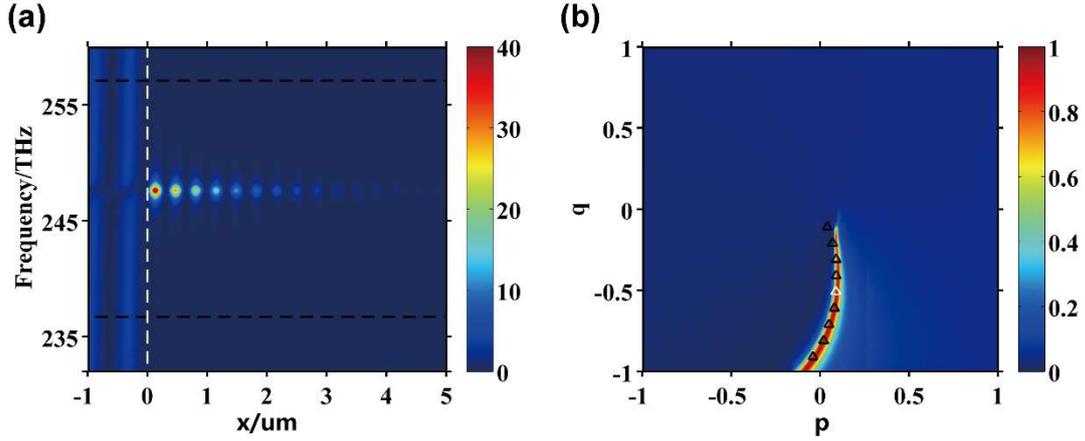

Figure S5 (a) the distribution of the intensity in the configuration consisting of a PC and a 15nm-thick silver film. The white dashed line indicates the interface between the PC and the silver film, and the black dashed lines label the edges of band gap of the PC. (b) the absorption of the same configuration in the p-q space with the frequency fixed at the frequency of the synthetic Weyl point. The black triangles mark the trajectory of the cavity mode resonances at this working frequency. In both (a) and (b), the parameters of the PC are $d_a = 97nm$, $d_b = 72nm$. The PC has 80 unit cells and is truncated at the center of the first layer. The relative permittivity of silver at the working frequency is $\varepsilon_{Ag} = -71.972 + 5.586i$. The p and q values for the PC in (a) are $p = 0.09$ and $q = -0.51$, which is also marked by a white triangle in (b).

# Reference

[1]   A. Dubois, J. Selb, L. Vabre, and A.-C. Boccara, *Phase measurements with wide-aperture interferometers*, App. Opt. **39**, 2326 (2000).

[2]   W. S. Gao, M. Xiao, C. T. Chan, and W. Y. Tam, *Determination of Zak phase by reflection phase in 1D photonic crystals*, Opt. Lett. **40**, 5259 (2015).

[3]   C. Fang, L. Lu, J. Liu, and L. Fu, *Topological semimetals with helicoid surface states*, Nat. Phys. **12**, 936 (2016).

[4]   K. J. Vahala, *Optical microcavities*, Nature **424**, 839 (2003).

[5]   http://refractiveindex.info.